\documentclass[trackchanges,twocolumn]{aastex7}
\usepackage{longtable}
\usepackage{placeins}
\usepackage{threeparttable}
\usepackage{scalerel}
\usepackage{amsmath}
\usepackage{float}
\usepackage{graphicx}
\usepackage{mathptmx}
\usepackage{txfonts}
\usepackage[T1]{fontenc}
\usepackage{ae,aecompl}
\newcommand\HI{H\protect\scaleto{$I$}{1.2ex}}

\newcommand{\kms}{\mbox{km\,s$^{-1}$}}

\usepackage{changepage}

\newcommand{\ttt}[1]{{\texttt{#1}}}

\newcommand*{\dittostraight}{---\textquotedbl---} 

\begin{document}
\title{Can dwarf spheroidal galaxies host a central black hole?}

\author[orcid=0000-0002-1250-4359,sname='K. Aditya']{K. Aditya}
\affiliation{Indian Institute of Astrophysics, Koramangala, Bengaluru 560 034, India}\email[show]{kaditya.astro@gmail.com}  
\author[orcid=0000-0001-9282-0011,sname='A. Mangalam']{A. Mangalam}
\affiliation{Indian Institute of Astrophysics, Koramangala, Bengaluru 560 034, India}
\affiliation{Faculty of Physical Sciences, Academy of Scientific and Innovative Research (AcSIR), India}
\email[show]{ mangalam@iiap.res.in}
\begin{abstract}
We construct mass models of Milky Way dwarf spheroidal galaxies to place constraints on the central black hole (BH) masses they can host.
We model the galaxies as a three-component system consisting of the stars, dark matter halo, and a central black hole, using the 
Osipkov–Merritt-Cuddeford class of anisotropic distribution function. The posterior distribution of black hole mass remains flat 
toward the low-mass end, indicating that the kinematic data places an upper limit on the black hole mass. Our analysis yields a 95\% credible upper limit of $\log(M_{\bullet}/M_{\odot}) < 6$. We combine our results with black hole mass measurements and upper limits from the literature to construct a unified $ M_{\bullet}\!-\!\sigma_{*}$ relation spanning $ \sigma_{*} \sim 10{-}300\,\mathrm{km\,s^{-1}}$, described by $\log(M_{\bullet}) = 8.32 + 4.08\,\log\left(\sigma_{*}/200\,\mathrm{km\,s^{-1}}\right)$, with an intrinsic scatter of $ \sigma_{\rm int} = 0.55$. We compare the inferred limits to models of black hole growth via momentum-driven accretion and stellar capture, which predict black hole masses in the range \( 10^{3}{-}10^{4}\, M_{\odot} \) for the range $ \sigma_{*} \sim 6{-}12 \,\mathrm{km\,s^{-1}}$, in close agreement with the \( M_{\bullet}\!-\!\sigma_{*} \) relation 
within the 95\% credible upper limits on the black hole masses derived in this work. 
\end{abstract}

\keywords{\uat{Galaxies}{573} --- \uat{Galaxy dynamics}{591} --- \uat{Galaxy kinematics}{602}  --- \uat{Dwarf spheroidal galaxies}{420} --- \uat{Intermediate-mass black holes}{816} --- \uat{Black holes}{162}}

\section{Introduction}
The dwarf spheroidal galaxies in the local group are simple 
dynamical systems and are ideal test beds for studying the 
structure and dynamics of dark matter halo. The dwarf spheroidal 
galaxies have relatively high stellar velocity dispersion 
\citep{walker2007velocity} compared to their luminosity 
($\rm M_{Dyn}/L_{V} > 10$) \citep{mateo1998dwarf, mcconnachie2012observed, battaglia2022stellar}, 
indicating that these galaxies are dominated by 
dark matter \citep{walker2014dark}. Across our sample, the central velocity dispersions lie in the range $6\kms$–$11\kms$ (median $8.5\kms$), while at the outermost measured radii, they range from $5\kms$ to $12.1\kms$ (median $7.5\kms$).
Further, the ratio of the \HI{} mass to the dynamical mass is equal to $\rm M_{\HI{}}/M_{Dyn}=10^{-5}$ 
\citep{grcevich2009h,spekkens2014dearth}, indicating that the 
dwarf spheroidal galaxies are extremely gas-poor. Thus, the 
relative simplicity of the dwarf spheroidal galaxies means that 
they can be modeled as a simple collisionless system consisting 
of \emph{stars + dark matter} without introducing the complexities 
brought about by the collisional gas component. Further, the dwarf 
spheroidal galaxies lie within the virial radius (<300kpc) of 
the Milky Way, providing detailed kinematics of the tracers and, 
thus, stringent constraints on the structure of dark matter halo.
The dwarf spheroidal galaxies have a common universal mass scale equal to $\rm M_{300}\simeq 10^{7}M_{\odot}$, within 300 parsecs, which  
sets the lower limit on the mass of the dark matter needed for 
galaxy formation \citep{strigari2008common,walker2009universal} 
making them ideal systems for studying galaxy formation at the smallest scales.

Recent studies have shown that, apart from the presence of dark matter and stars, 
dwarf spheroidal galaxies may also potentially host black holes at their centers. 
In a recent study, \cite{bustamante2021dynamical} show that the dwarf spheroidal galaxy 
Leo-1 hosts a central black hole with a mass equal to $\rm (3.3 \pm 2.0)\times10^{6}M_{\odot}$. 
Also, see \cite{pascale2024central}, who show that kinematics in the central region 
may only impose an upper limit equal to $\rm 10^{5}M_{\odot}$. Also, interestingly, the black 
hole mass estimated by \cite{bustamante2021dynamical} for Leo-1 exceeds the values predicted 
by the standard $\rm M_{\bullet}-\sigma_{*}$ relation \citep{kormendy2013coevolution}. 
Further, using deep radio observations \cite{maccarone2005upper} find the signature of 
$\rm 10^{4}M_{\odot}$ black hole in Ursa Minor (UM) dwarf spheroidal galaxy, also see 
\cite{lora2009upper}. In another study using deep XMM Newton observations \cite{manni2015xmm} 
find an upper limit on the black hole mass equal to  $\rm 2.3 \times 10^{6}M_{\odot}$ 
in UM. \cite{jardel2012dark} using Schwarzschild orbit superposition models, find an 
upper limit on black hole mass in another dwarf spheroidal galaxy, Fornax, equal to 
$\rm 3.2 \times 10^{4} M_{\odot}$. For the remaining dwarf spheroidal galaxies 
in our sample, Carina, Draco, Leo II, Sculptor, and Sextans, no direct measurements or upper 
limits on central black hole masses have been reported in the literature. Further studies by
\cite{reines2013dwarf,bellovary2019multimessenger,reines2022hunting} provide compelling 
observational evidence for black holes in the low-mass dwarf galaxies. 

In this paper, we construct dynamical models of dwarf spheroidal 
galaxies as \emph{stars + dark matter + black hole} and constrain 
the black hole mass in conjunction with the parameters corresponding 
to the dark matter halo. We model the dwarf spheroidal galaxies using 
the Osipkov–Merritt-Cuddeford class of anisotropic distribution functions 
\citep{osipkov1979spherical,merritt1985spherical,cuddeford1991analytic}. Distribution 
function–based models have been extensively employed to derive dynamical constraints on 
the properties of dark matter halos and central black holes. For spherical distribution 
function models, see \cite{posti2015action, pascale2019action, shchelkanova2021n, aditya2024challenges}, 
and for axisymmetric distribution-function models of galaxies, see \cite{aditya2021cold, aditya2022h, aditya2023h}.
We use a publicly available stellar dynamics toolbox called Action-based Galaxy 
Modelling Architecture (AGAMA) \citep{vasiliev2019agama} to compute the distribution 
function and other associated quantities like the line-of-sight velocity dispersion. 
We will derive the constraints on the black hole masses and inspect possible 
formation scenarios for black holes in dwarf spheroidal galaxies. Finally, we will 
discuss the estimated black hole masses in the context of unified $\rm M_{\bullet}-\sigma_{*}$ 
relation spanning $ \sigma_{*} \sim 10{-}300\,\mathrm{km\,s^{-1}}$. The $M_{\bullet}-\sigma_{*}$ relation in the  
low-$\sigma_{*}$ regime is particularly valuable for current simulations and future searches for 
intermediate-mass black holes \citep{nguyen2025simulating, ngo2025detecting} using instruments such as LIGER \citep{wright2024liger} on Keck or HARMONI on the upcoming 
39-meter Extremely Large Telescope (ELT) \citep{thatte2022harmoni}.

The paper is organized as follows: in \S 2, we introduce the dynamical models of the dwarf spheroidal galaxies and present the results and discuss their implications in \S 3. Finally, we will draw our conclusions in \S4.

\section{Dynamical model of dwarf spheroidal galaxy}

\begin{table*} 
\begin{adjustwidth}{-1.5cm}{0cm}
\caption{Observational data corresponding to the dwarf spheroidal galaxies.}
\resizebox{18cm}{!}{\begin{tabular}{|l|c|c|c|c|c|c|c|}	
\hline
Galaxy             & RA(J2000)      &   Dec(J2000) &   Distance   &  $\rm M_{*}$             & $\rm R_{*}$  &References  &References \\
                   & $\rm (hh:mm:ss)$     &   $\rm (dd:mm:ss)$ &   $\rm (kpc)$      &  $\rm (10^{6}M_{\odot})$   & $\rm (pc)$ & Photometry & Velocity dispersion           \\
\hline       
\hline
Carina             & $06:41:36.7$  & $-50:57:58$ & $106 \pm 6$  &  0.38   &  $308 \pm 23$  & \cite{munoz2018megacam},&\cite{walker2007velocity}  \\
Draco              & $17:20:12.4$  & $+57:54:55$ & $76  \pm 6$  &  0.29   &  $214 \pm 2$   & \dittostraight&\dittostraight  \\
Fornax             & $02:39:59.3$  & $-34:26:57$ & $147 \pm 12$ &  20     &  $838 \pm 3$   & \dittostraight&\dittostraight   \\
Leo-1              & $10:08:28.1$  & $+12:18:23$ & $254 \pm 15$ &  5.5    &  $270 \pm 2$   & \dittostraight&\dittostraight   \\
Leo-2              & $11:13:28.8$  & $+22:09:06$ & $233 \pm 14$ &  0.74   &  $171 \pm 2$   & \dittostraight&\dittostraight  \\  
Sculptor           & $01:00:09.4$  & $-33:42:33$ & $86  \pm 6$  &  2.3    &  $280 \pm 1$   & \dittostraight&\dittostraight  \\
Sextans            & $10:13:03.0$  & $-01:36:53$ & $86  \pm 4$  &  0.44   &  $413 \pm 3$   & \dittostraight&\dittostraight   \\
UM                 & $15:08:08.5$  & $+67:13:21$ & $76  \pm 3$  &  0.29   &  $407 \pm 2$   & \dittostraight&\cite{walker2009universal}  \\
\hline
\end{tabular}}
\label{table: table 1}
\end{adjustwidth}
\end{table*}

We model the dwarf spheroidal galaxy as a collisionless system of stars and dark matter with a central black hole. We use the observed stellar photometry \citep{munoz2018megacam} of dwarf spheroidal galaxies for constructing the potential of the stellar distribution. Fixing the stellar density from observations, alleviates the degeneracy between the stellar and the dark matter distribution, allowing us to place better constraints on the dark matter density 
profile parameters and the black hole mass. We adopt the structural parameters of the dwarf spheroidal galaxies from \citet{munoz2018megacam}, which were measured using a Plummer density profile. In our model, we keep the parameters corresponding to the potential of the dark matter halo and the black hole masses as free parameters. Once we construct the total potential, we numerically compute the stellar distribution function in the total potential and constrain our model parameters using the observed stellar velocity dispersion \citep{walker2007velocity,walker2009universal}. We estimate the posterior distribution of our model parameters using \textit{emcee} \citep{foreman2013emcee}.

\subsection{Stellar density}
We model the stellar density using the Plummer potential \citep{plummer1911problem}  given by;
\begin{equation}
\Phi(r) = -\frac{GM_{*}}{\sqrt{R^2 + R_{*}^2}}.    
\end{equation}
The density corresponding to this potential is given by;
\begin{equation}
\rho(r) = \frac{3M_{*}}{4\pi R^3_{*}} \left(1 + \frac{R^2}{R_{*}^2}\right)^{-\frac{5}{2}}.  
\end{equation}
In the above equations, $R_{*}$ is the scalelength of the stellar density and  $M_{*}$ is the mass of the stellar density profile. The parameters corresponding to the stellar density profile of the Milky Way dwarf spheroidal galaxies studied in the current work are taken from 
\citep{munoz2018megacam} and are summarized in Table 1.

\subsection{Dark matter density}
We model the dark matter density using Navarro Frenk White (NFW) profile \citep{navarro1997universal} given by;
\begin{equation}
\Phi_{DM}(R) = -4 \pi G \rho_{DM} R_{s}^3 \frac{\ln\left(1 + \frac{R}{ R_{s} }\right)}{R}.
\end{equation}
The corresponding  density profile is given by;
\begin{equation}
 \rho(r) = \frac{\rho_{DM}  }{\frac{R}{R_{s}  } \left(1 + \frac{R}{R_{s} }\right)^2}.   
\end{equation}
In the above equation, $\rho_{DM}$ and $R_{s}$ are the density and scalelength of the dark matter density profile 
respectively. 

\subsection{Black hole }
We model the black hole as,
\begin{equation}
    \Phi_{\bullet}=\frac{GM_{\bullet}}{R},
\end{equation}
where $M_{\bullet}$ is the mass of the black hole.
\subsection{Galaxy Model}
We combine the density of the stellar component and dark matter models to compute the total potential of 
the system using the \texttt{AGAMA} class \texttt{Potential}. The \texttt{Potential} class in \texttt{AGAMA} uses multipole expansion to compute the potential for any given density profile. The density is decomposed into spherical harmonics with radially 
varying amplitudes. Finally, the potential corresponding to each spherical harmonics term is combined
to compute the total potential (see \cite{vasiliev2019agama}). Once the density is defined, the potential can be computed by calling the built-in Poisson solver implemented in \ttt{AGAMA} class \ttt{Potential}. We define the generalized anisotropic distribution function \citep{osipkov1979spherical,merritt1985spherical,cuddeford1991analytic} given 
by;
\begin{equation}  
f(E,L) = \hat f(Q) \; L^{-2\beta_0}, \qquad Q = E - L^2 / (2 R_a^2),
\end{equation}
where $R_{a}$ is the anisotropy radius. In the above formula $E=\Psi_{T}- v^{2}/2$ is 
the binding energy, $\Psi_{T}=-\phi_{T}$ is the relative total potential due to the combined effect of all the density components, and $f(Q)=0\, \forall\, Q\leq0$.
The term $L^{-2\beta_{0}}$ regulates the orbital anisotropy. When $\beta_{0}=0$, the distribution function is ergodic, and the orbits are isotropic, 
whereas $\beta_{0}<0$ and $\beta_{0}>0$ produce distribution functions which describe tangentially and radially biased orbits, respectively. 
In Osipkov–Merritt-Cuddeford class of anisotropic distribution function, the density component depends on the energy and on the angular 
momentum modulus of stellar orbits through the variable $Q$. The $Q$ parameter controls the radial variation of anisotropy. The 
distribution function is ergodic within the anisotropy radius $(r<<R_{a})$  and becomes radially biased beyond the anisotropy radius $r>>R_{a}$. The parameters of our dynamical model are described in Table 2.
The distribution function produces a velocity anisotropy 
\begin{align}
\beta(r)\equiv 1 - \frac{\sigma_T^2}{2\sigma_R^2} =  \frac{ R^2  +\beta_0 R^{2}_a }{R^{2}_a + R^2},
\end{align}
where $\rm \sigma_T^2$ and $\rm \sigma_R^2$ are the tangential and the radial velocity dispersion, respectively.
In the above equation $\rm \beta_0$ is the value of anisotropy at the center and $\rm R_{a}$ is 
the anisotropy radius. The isotropic case is obtained by setting $\rm \beta_0=0, \, R_a=\infty$. $\rm \hat f(Q)$ 
is obtained through a more general Eddington inversion formula derived by \cite{osipkov1979spherical,
merritt1985spherical,cuddeford1991analytic,ciotti2021introduction}. The distribution function is 
tangentially biased when $\beta_{0}<0$, radially biased when $\beta_{0}>0$, and isotropic when $\beta_{0}=0$.
The distribution function $\rm \hat f(Q)$ is given by \citep{cuddeford1991analytic,ciotti2021introduction},
\begin{align}
\hat f(Q) &= \frac{(-1)^{m+1}\cos(\beta_{0}\pi)}{\pi\, 2^{\beta_{0}} (2\pi)^{3/2}} 
\cdot \frac{\Gamma\left(\frac{3}{2} + \beta_{0} - m \right)}{\Gamma(1 + \beta_{0})} \cdot \frac{d}{dQ} \\
&\quad \times \int^{Q}_{0} \frac{d^{m} \hat \rho }{d\Psi_{T}^{m}} 
\cdot \frac{d\Psi_{T}}{(Q - \Psi_{T})^{\frac{3}{2} + \beta_{0} - m}}
\end{align}

where, $\rm m$ is defined as 
\begin{align}
m \equiv \rm int( \beta_{0} + \frac{1}{2} ) +1 .    
\end{align}
We fix the density of the stellar component directly from observations and calculate the corresponding 
potential by solving the Poisson's equation using multipole expansion methods implemented in \texttt{AGAMA} 
class \texttt{Potential}. We then combine the potential due to the stellar component with the potential corresponding 
to the dark matter halo and the black hole. The parameters corresponding to the black hole and the dark matter halo are 
treated as free parameters in our model, whereas the potential for the stellar component is fixed from the observations. 
Once the potentials and density for each component are self-consistently defined by solving the Poisson's equation in \texttt{AGAMA}, 
the distribution function that generates the given density is computed using the \ttt{QuasiSpherical} model implemented in 
the \ttt{DistributionFunction} class in \ttt{AGAMA}. The \ttt{DistributionFunction} class numerically computes the augmented density $\hat \rho(\Psi_{T})$ in terms of the potential and computes the corresponding distribution function. In the above equation, $\Psi_{T}$ is the total relative potential of the system given by $\Psi_{T}=-\phi_{T}= \phi_{*} +\phi_{DM} + \phi_{\bullet}$. Finally, we combine the density and the distribution function using the \ttt{AGAMA} module \ttt{GalaxyModel} and use the task \ttt{moments} to compute the line-of-sight velocity dispersion as a function of radius. We compare the modeled velocity dispersion to the observed velocity dispersion using $\rm \chi^{2}$ defined as
\begin{equation}
\chi^{2} =\sum _{R} \frac{\bigg(\sigma_{\rm{obs}}(R) - \sigma_{\rm model}(R) \bigg)^{2} }{s^{2}_{\rm{err}}(R)},
\end{equation}
where, $\rm \sigma_{\rm{obs}}$ is the observed stellar dispersion, $\rm \sigma_{\rm{{model} }}$ is the modeled line of sight dispersion and $\rm s^{2}_{err}$ is the error on the observed dispersion.

\begin{table}
    \centering
    \begin{tabular}{|c|c|}
        \hline
        Symbol & Description  \\
         \hline
         \hline
         $\rm M_{\bullet}$                          &Black hole mass \\
         $\rm M_{DM},\, \Phi_{DM},\, \rho_{DM} $    &Mass, potential, and \\ 
                                                    &density of the dark matter halo\\
         $\rm M_{*},\, \Phi_{*},\, \rho_{*} $       &Mass, potential, and the \\
                                                    &density of stars\\
         $\phi_{T}= \phi_{\star}+ \phi_{DM} + \phi_{\bullet}$    &Total potential\\ 
         $\rm \Psi_{T}=-\phi_{T}$                   &Relative total potential\\
         $\rm R_{s}$                                &Scale radius of dark matter\\ 
                                                    &density profile  \\
         $\rm R_{*}$                                &Scale radius of stellar density profile\\ 
         $\rm \beta_{0}$                            &Anisotropy at the center \\    
         $\rm R_{a}$                                &Anisotropy radius\\
                     \hline        
    \end{tabular}
    \caption{Parameters describing the dynamical model of a dwarf spheroidal galaxy.}
    \label{tab:simple_table1}
\end{table}
\section{Results and Discussion}

\begin{figure*}
\tabcolsep=0.1\linewidth
 \divide\tabcolsep by 8 
 \begin{tabular}{cc}
    \includegraphics[width=0.50\linewidth]{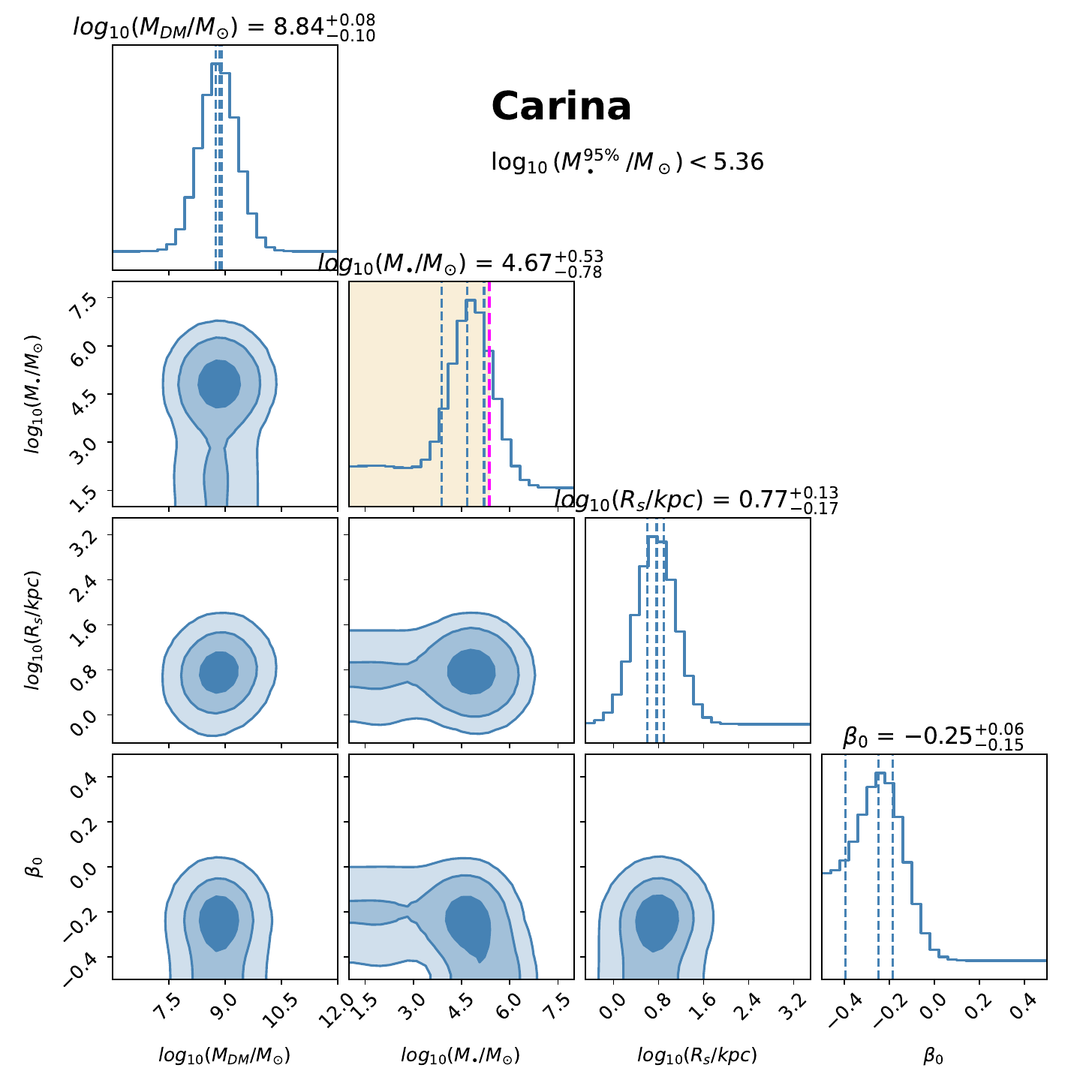} &
    \includegraphics[width=0.50\linewidth]{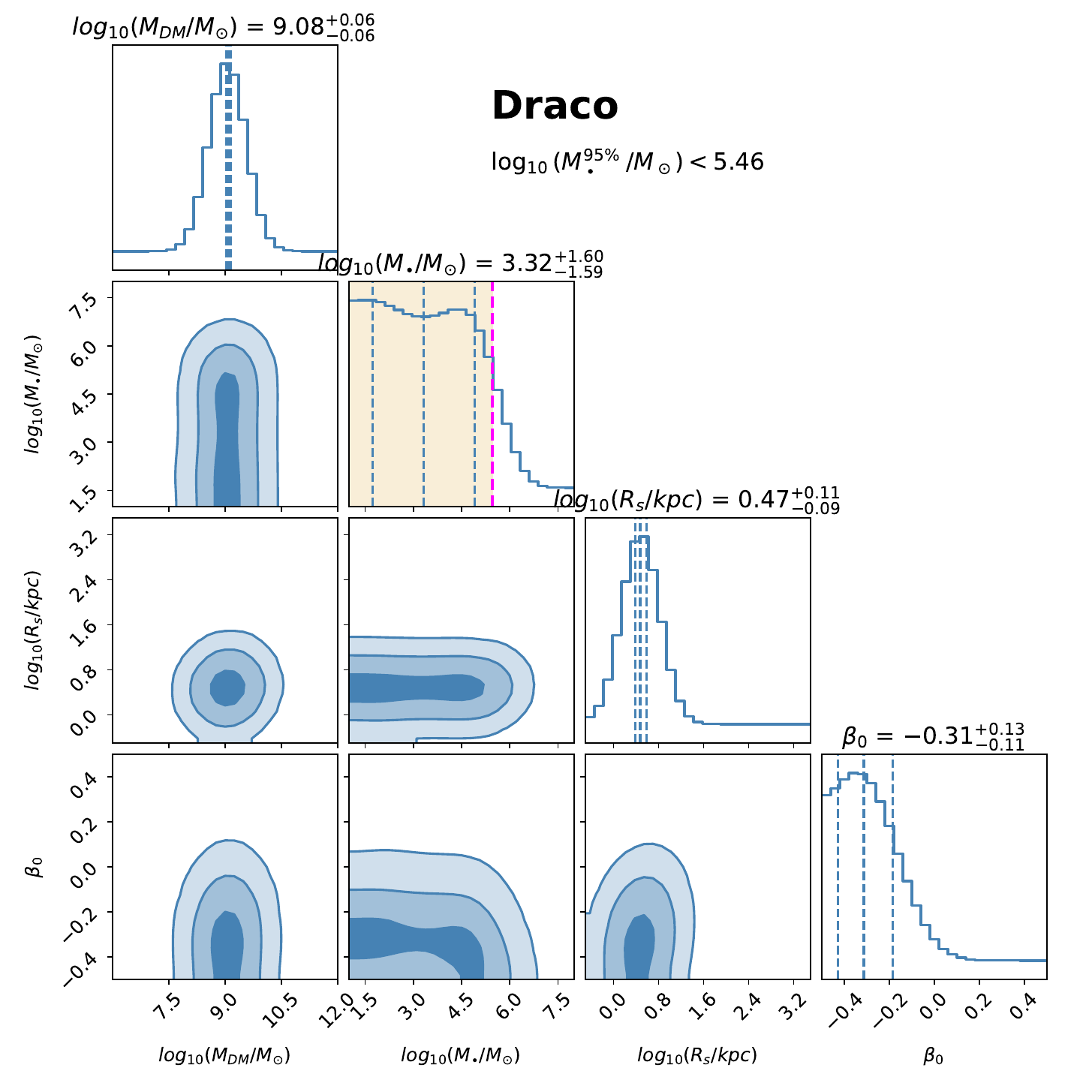} \\
    \includegraphics[width=0.50\linewidth]{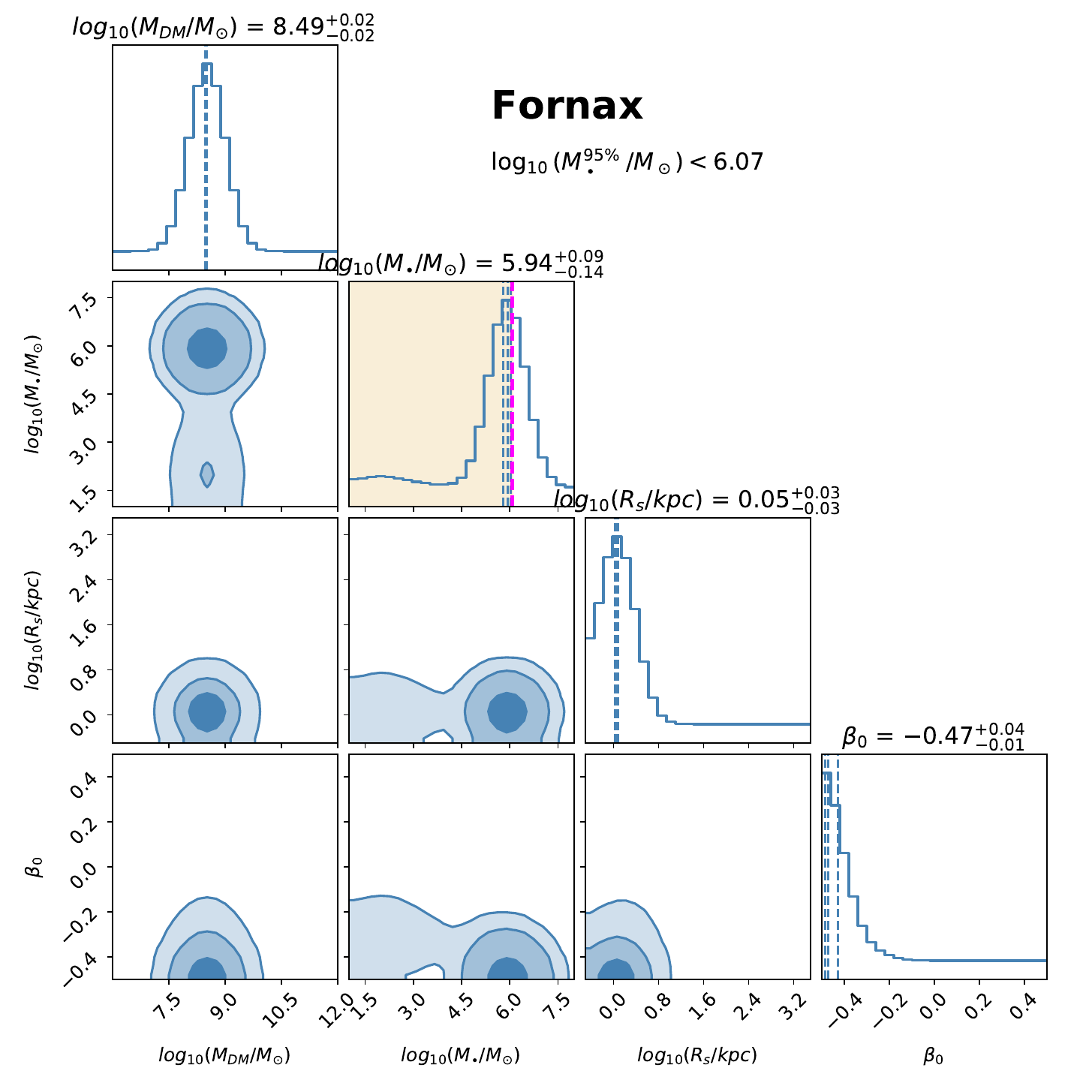} &
    \includegraphics[width=0.50\linewidth]{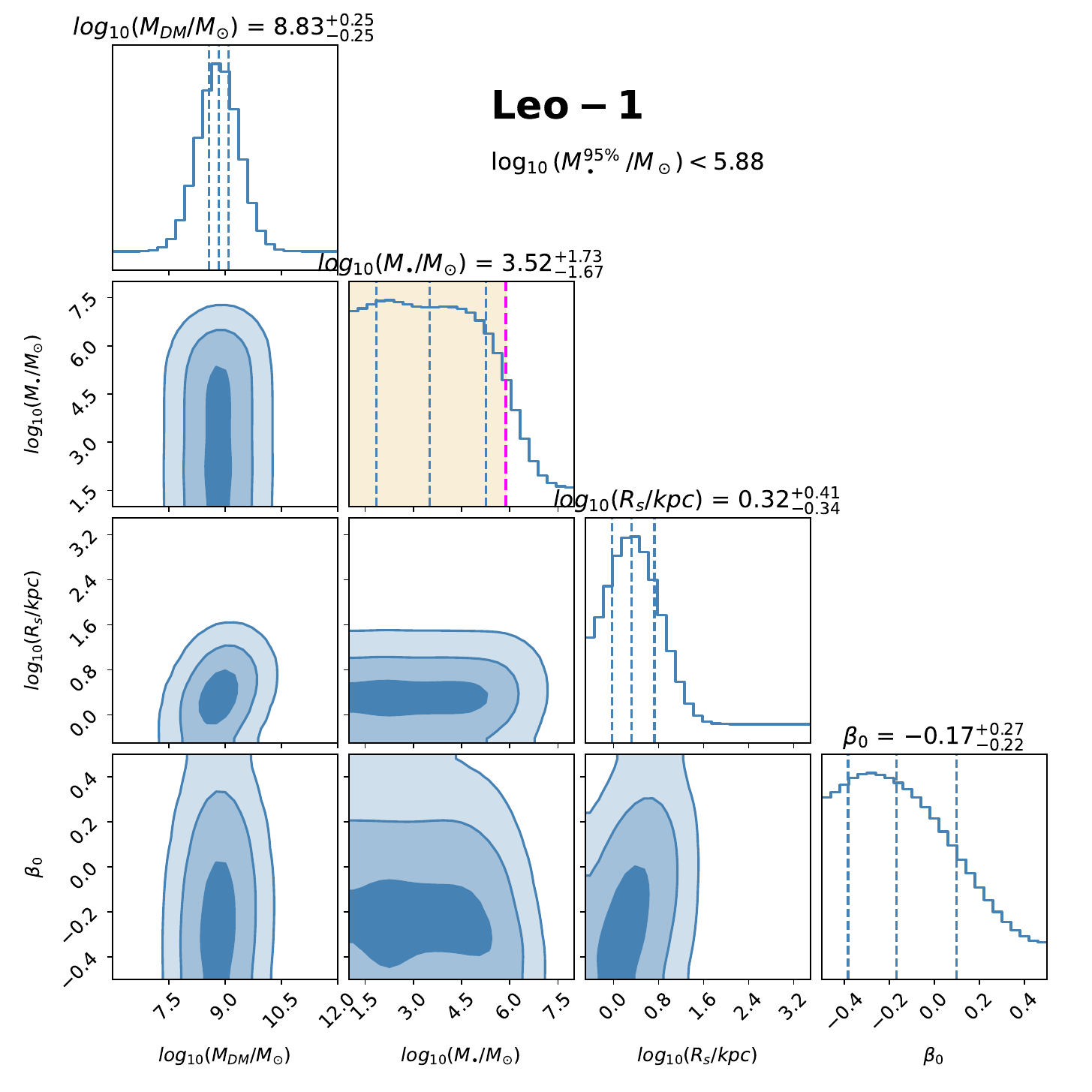} 
  \end{tabular}
\caption{Posterior probability distribution corresponding to model parameters of dwarf spheroidal galaxies; Carina, Draco, Fornax, and Leo-1. 
The dashed blue lines depict the $16^{th}$, $50^{th}$, and $84^{th}$ percentiles of the posterior probability distribution. 
The posterior distribution for the black hole mass is flat towards the low-mass end and drops off at $95\%$ 
credible upper limit of $\log(M_{\bullet}/M_\odot) < 6$, as indicated by the vertical magenta line.}  
\end{figure*}

\begin{figure*}
\tabcolsep=0.1\linewidth
 \divide\tabcolsep by 8
 \begin{tabular}{cc}
    \includegraphics[width=0.50\linewidth]{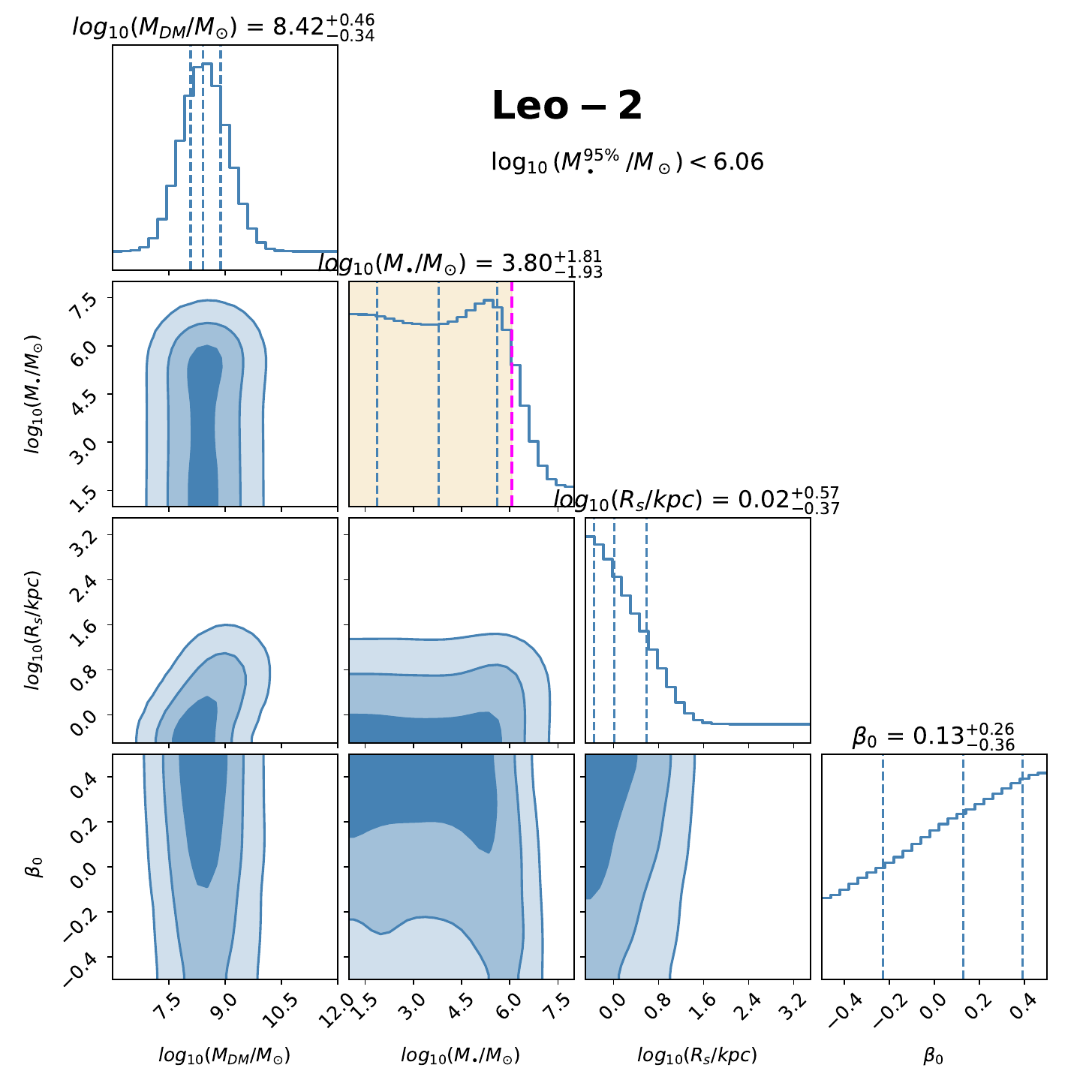} &
    \includegraphics[width=0.50\linewidth]{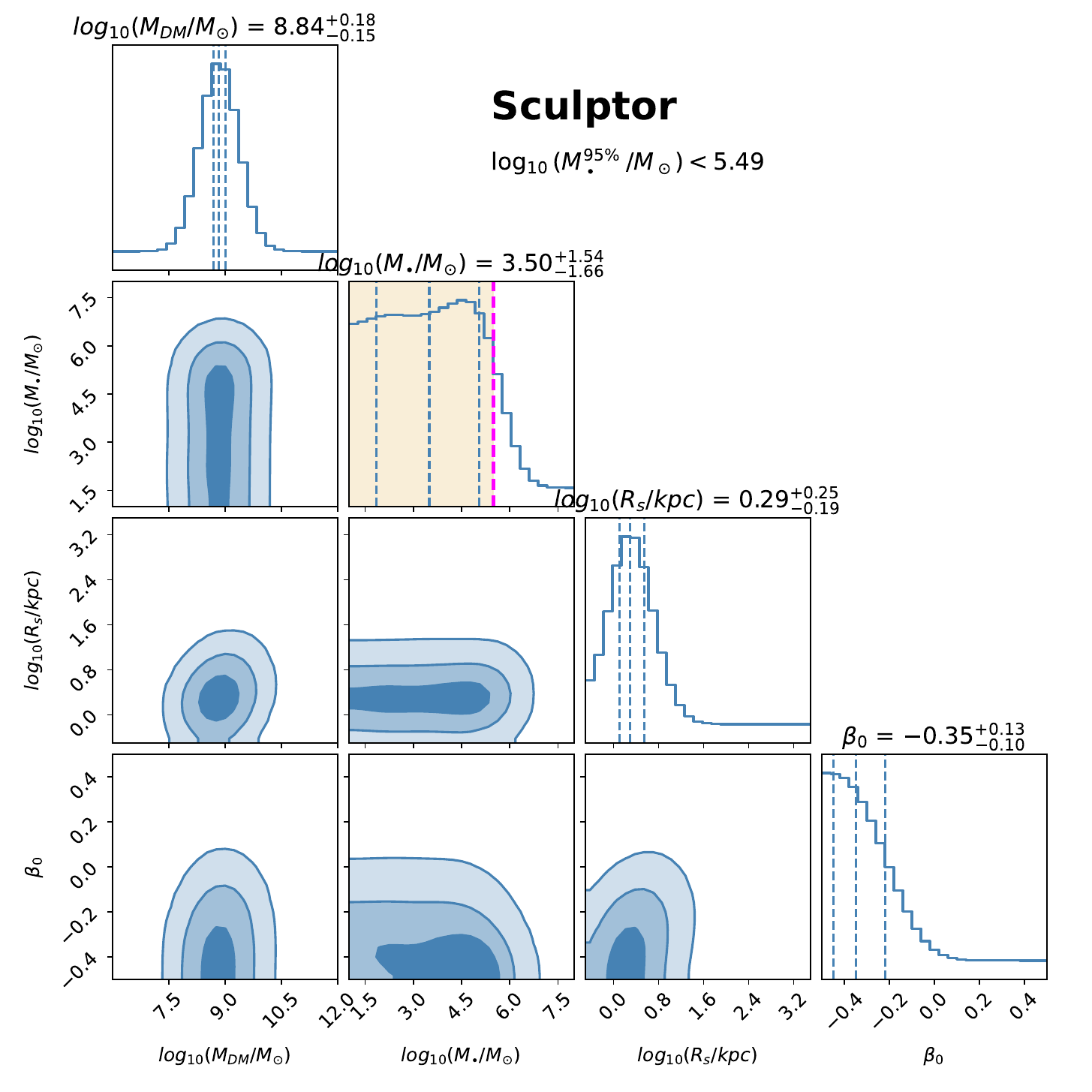} \\
    \includegraphics[width=0.50\linewidth]{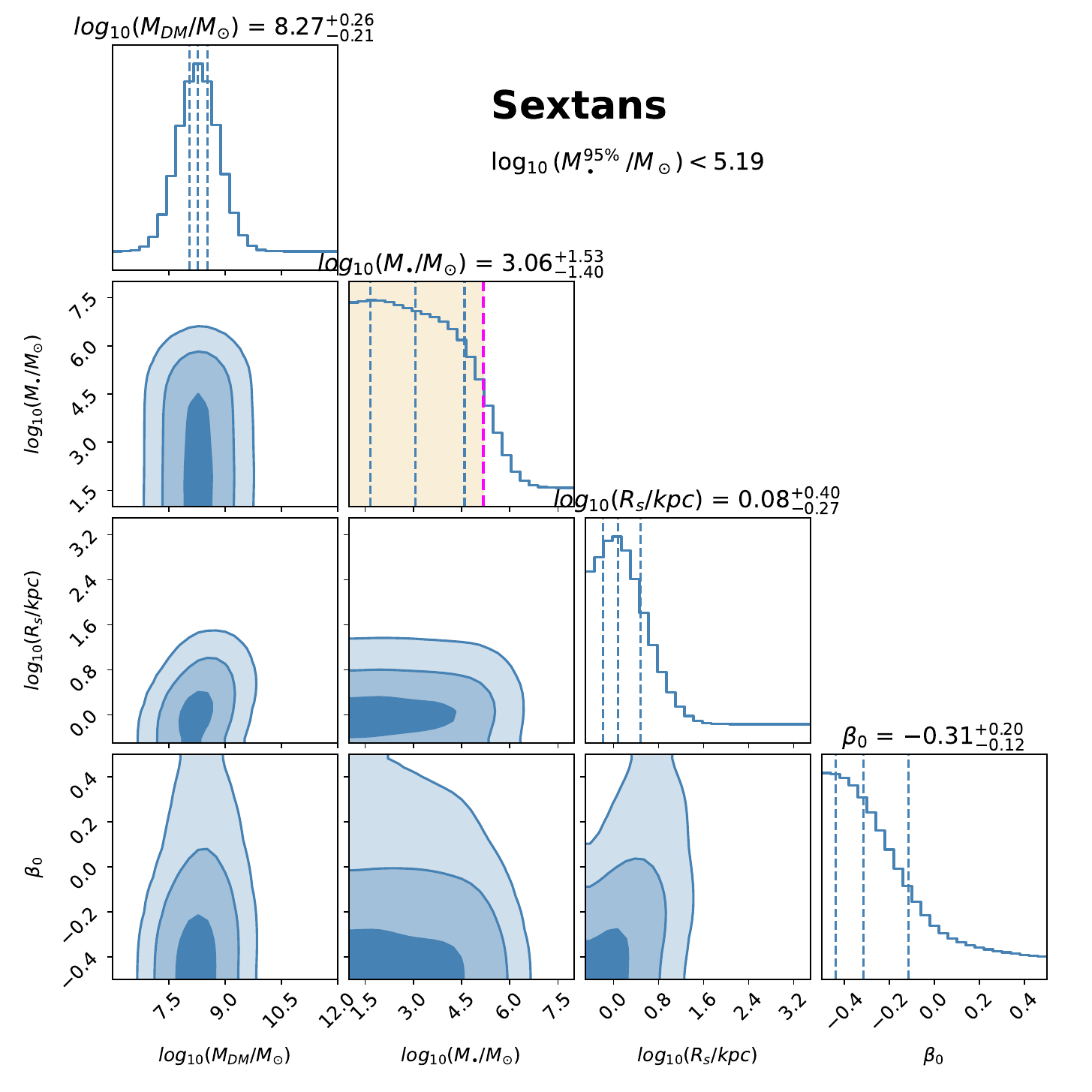} &
    \includegraphics[width=0.50\linewidth]{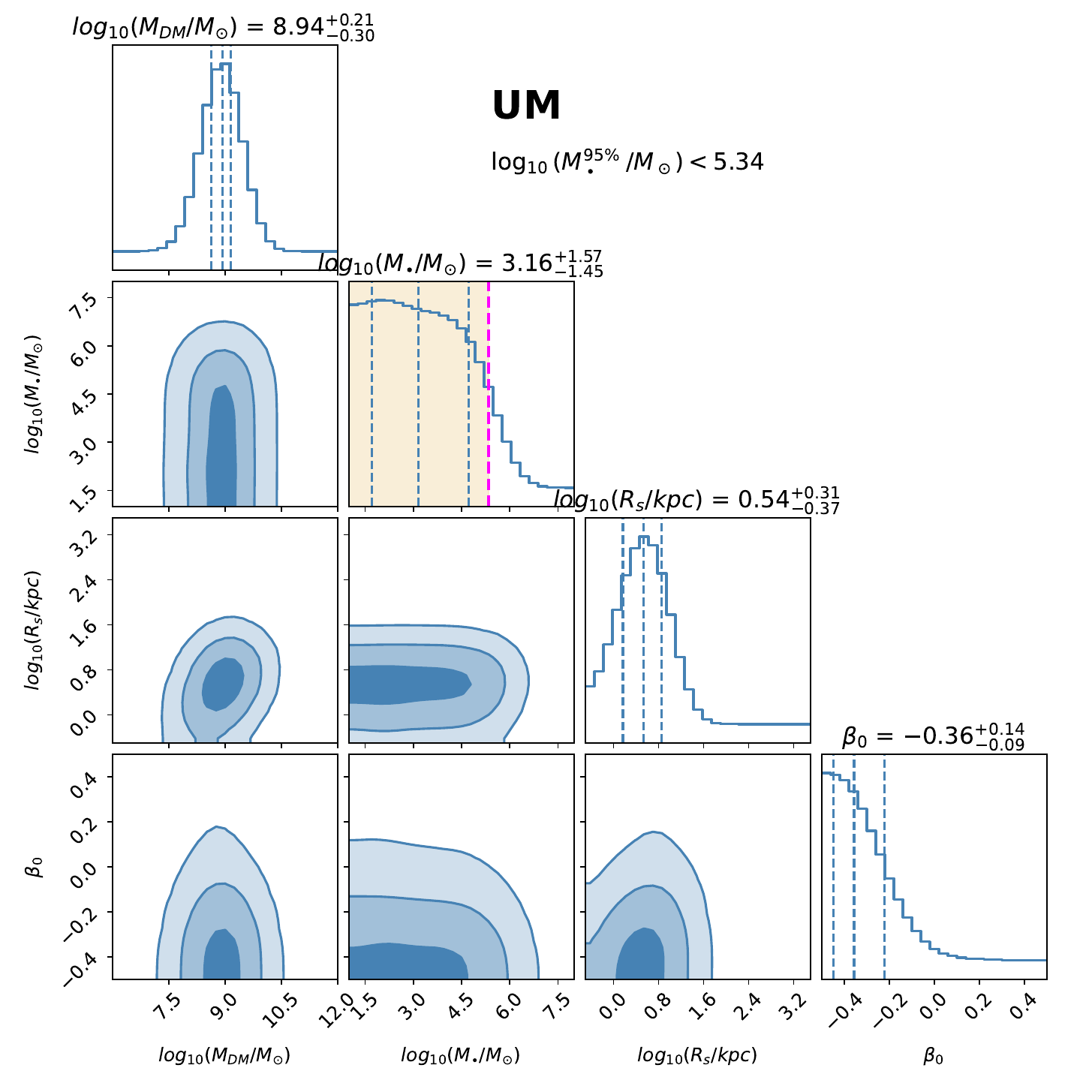} 
  \end{tabular}
\caption{Same as Figure 1, but for the dwarf spheroidal galaxies Leo-2, Sculptor, Sextans, and UM.} 
\end{figure*}

\begin{figure*}
\hspace*{3mm}
\resizebox{170mm}{80mm}{\includegraphics{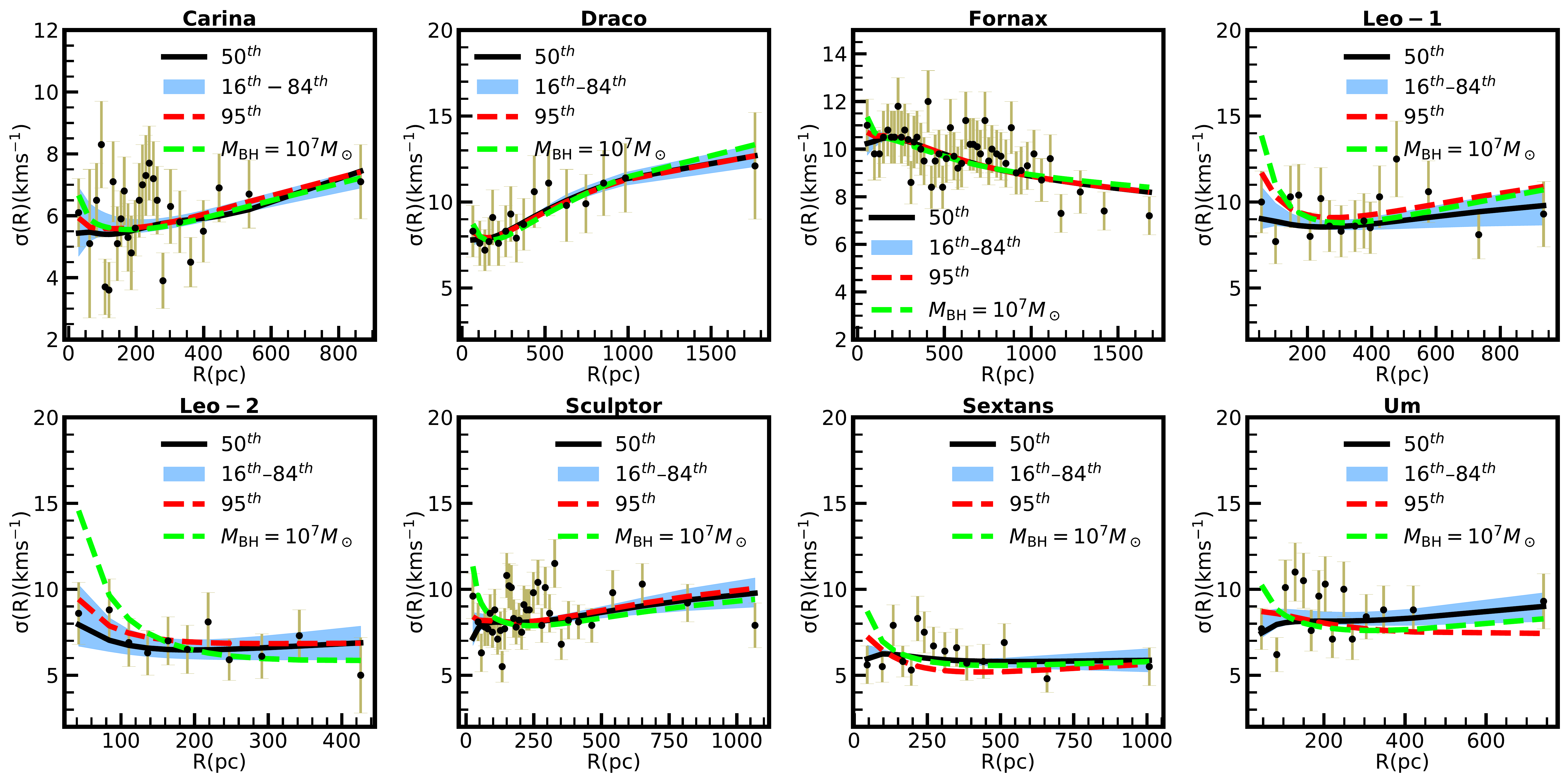}} 
\caption{The black line indicates the 50$^{\mathrm{th}}$ percentile, and the red curve represents the $95\%$ 
credible upper limit of the posterior distribution. The blue shaded region indicates the 16$^{\mathrm{th}}$-84$^{\mathrm{th}}$ percentile range. 
The green curve represents a model with $M_{\bullet} = 10^{7}\,M_{\odot}$, selected from the MCMC posterior samples.}
\end{figure*}

We computed the posterior probability distribution of the model parameters using the "emcee" Markov Chain
Monte Carlo (MCMC) ensemble sampler \citep{foreman2013emcee}. We adopted uniform priors on the model parameter:

\begin{enumerate}
\item $6.0$            <  $\log(M_{DM}/M_{\odot})$          <  $10$\\
\item $1.0$            <  $\log(M_{\odot}/M_{\odot})$       <  $7$\\
\item $-0.5$           < $\log(R_{s}/kpc)$                  <  $1.0$\\
\item $-0.5$           < $\beta_{0}$                        <  $0.5$\\
\end{enumerate}

The above priors on the model allow us to explore the posterior distribution of dark matter halos $(M_{DM},\, R_{s})$ and 
black hole masses  $(M_{\bullet})$ over a large parameter range. We sample $M_{DM}$, $R_{s}$ and 
$M_{\bullet}$ in logarithmic space to ensure uniform sampling of parameters that span several 
orders of magnitude. Furthermore, log uniform priors assign equal weight to samples per decade 
and produce a nearly Gaussian posterior distribution. The range of priors also encompasses the stellar distributions
with tangentially biased anisotropy $(\beta_{0}<0)$, a radially biased anisotropy profile $(\beta_{0}>0)$, and isotropic models 
$(\beta_{0}=0,\, R_{a} \rightarrow \infty)$. The bound on $\beta_{0}\geq-0.5$ is imposed by the implementation of 
the Osipkov–Merritt-Cuddeford class of anisotropic distribution function in \texttt{AGAMA}. When $\beta_{0}<-0.5$, the 
derivative of the augmented density $(\hat \rho(\Psi_{T}))$ becomes poorly conditioned at the center, and the distribution function 
becomes numerically unstable. So models implemented in \texttt{AGAMA} are restricted to $\beta_{0}>-0.5$, which ensures that the
distribution function is physical and numerically stable. Our preliminary fits showed that the inferred values of $R_{a}$ exceeded the physical extent of the galaxy, indicating that the models exhibit constant velocity anisotropy within the kinematic extent of the galaxy. So, we fix the value of $R_{a} = \infty$. We show the posterior distribution of our model parameters in Figures 1 and 2, respectively. The posterior for the dark matter mass is strongly constrained, with the median value of $\log(M_{\mathrm{DM}}/M_{\odot})$ lying between 8.3 and 9.1. The dark matter masses for the dwarf spheroidal galaxies are consistent with $\log(M_{300}/M_{\odot}) = 7$, as estimated by \citet{strigari2008common, walker2009universal, walker2014dark}, and with $\rho_{\mathrm{DM}}(150)\,/\,M_{\odot}\,\mathrm{kpc}^{-3} = (7$--$20)\times10^{7}$, in agreement with \citet{read2019dark, hayashi2020diversity}. The posterior distribution for the dark matter scalelength is also well constrained, with values lying in the range $\log(R_{s}/\mathrm{kpc}) = 0.02$-$0.8$. The posterior for the black hole mass remains flat toward the low-mass end, indicating that the kinematic data only place upper limits on the black hole mass. Therefore, we quote the 95\% credible upper limit instead of the median. Our analysis yields 95\% credible upper limits in the range $\log(M_{\bullet}/M_{\odot}) < 5.1$--$6.1$ for black hole masses for our sample of dwarf spheroidal galaxies. We overlay the model velocity dispersion profiles corresponding to the 50$^{\mathrm{th}}$ and 95$^{\mathrm{th}}$ percentiles of the posterior distribution on the observed velocity dispersion data in Figure 3. For comparison, we select a model with $M_{\bullet} =10^{7}\,M_{\odot}$ from the MCMC chain and overlay its predicted profile on the data. The kinematic data constrain the black hole mass to be below $10^{6}\,M_{\odot}$ within the observed uncertainties, as models with $10^{7}\,M_{\odot}$ clearly overpredict the velocity dispersion. The models are tangentially biased, with $\beta_0 < 0$. However, we note that $\beta_{0}$ is unconstrained for Leo-2, likely due to insufficient kinematic data close to the center. The posterior distribution for black hole masses indicates that the dwarf spheroidal galaxies around the Milky Way can at most accommodate an intermediate mass black hole (IMBH).

\subsection{Formation scenarios for IMBH}
 It has been shown  that the black 
holes at the center of galaxies grow by accreting gas \citep{1982MNRAS.200..115S,tremaine2002slope}. 
\cite{king2003black} shows that $M_{\bullet} - \sigma_{*}$ 
relation emerges naturally when the outflows from the black hole
can communicate with the ambient medium through exchange 
of momentum and drive a significant outward flow. If the 
black hole mass is greater than the ambient medium, and the 
surrounding gas is completely expelled, then the saturation  
black hole mass \citep{merritt2013dynamics} is given by
\begin{equation}
    M_{\bullet} =2 \times 10^{8}\left(\frac{f_{g}}{0.1}\right) \left(\frac{\sigma_{*}}{200\kms}\right)^{4} M_{\odot},
\end{equation}
where $f_{g}$ is the gas fraction. The black hole mass 
saturates to the values predicted by the $\sigma^{4}$ law 
when the outflow velocity exceeds the escape velocity of
the medium, driving away the ambient gas and halting the accretion
process. Using the value of the cosmic baryon 
fraction $f_{g}=0.16$ \citep{2009ApJS..180..330K} and typical 
velocity dispersion for our sample of galaxies in (12), we obtain 
$M_{\bullet}\approx 10^{3}M_{\odot}$. The saturation mass typically sets 
the upper limits on the seed black hole mass in dwarf spheroidal 
galaxies $\leq 10^{3}M_{\odot}$. The critical mass below which the 
stellar capture dominates over accretion \citep{bhattacharyya2020cosmic} is given by 
\begin{equation}
    M_{c}= 5\times10^{3} \eta^{-0.75}M_{\odot},
\end{equation}
where $\eta=\dot M_{\odot}/\dot M_{E}$, $\dot M_{E}$ is the Eddington accretion rate.  Assuming sub-Eddington accretion $\eta\geq 0.07$, we get critical mass $\approx 10^{4} M_{\odot}$, higher than the saturated black hole mass. Thus, stellar capture may be an important channel through which the black hole can grow in dwarf spheroidal galaxies. It is essential to note that the growth of a black hole due to accretion is halted when the ambient medium is completely expelled. However, the black hole continues to grow by accreting mass, capturing stars, and potentially reaching masses of $ 10^{4}\, M_{\odot}$ and above as shown in \cite{bhattacharyya2020cosmic} (see equation (74) and Fig. 17 for estimates).

\subsection{Unifying the $\rm M_{\bullet} - \sigma_{*}$ Relation}
\begin{figure*}
\hspace{-1.5cm}
\resizebox{220mm}{140mm}{\includegraphics{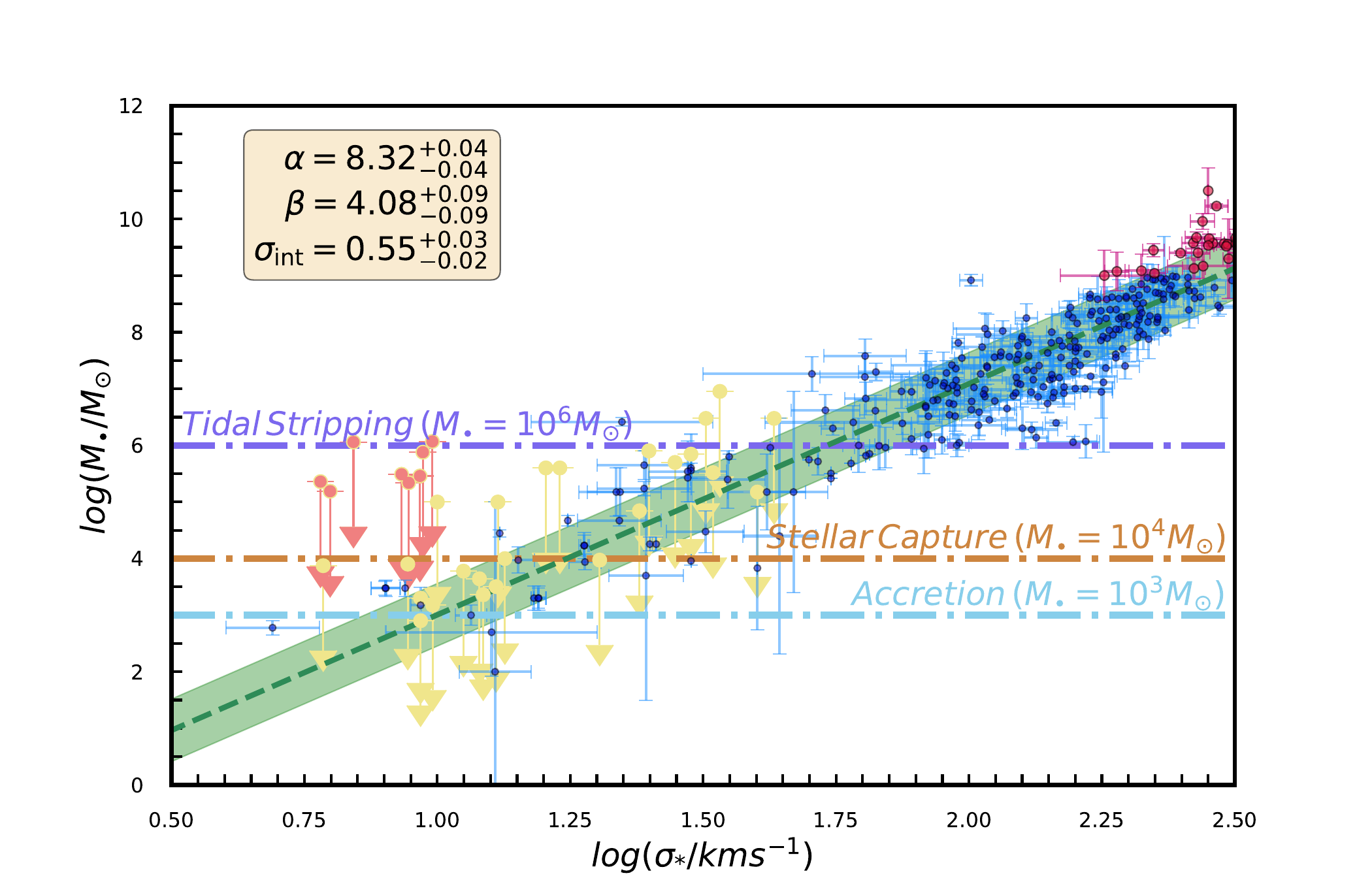} }
\caption{A unified $\rm M_{\bullet}$--$\sigma_{*}$ relation spanning stellar velocity dispersions from $\sim 10~\mathrm{km\,s^{-1}}$ to $\sim 300~\mathrm{km\,s^{-1}}$. 
Blue points represent black hole mass estimates, while yellow arrows denote upper limits, compiled from \protect\cite{lutzgendorf2013m, van2016unification, greene2020intermediate}. The red arrows indicate the upper limits obtained in the present work. The green line shows the best-fit regression, and the shaded region indicates $1\sigma$ scatter. Magenta points depict ultra massive black holes $(M_{\bullet}>10^{9}$. Theoretical limits on black hole masses (for our sample range $ \sigma_{*} \sim 6{-}12 \,\mathrm{km\,s^{-1}}$)  due to accretion, stellar capture, and tidal stripping are also overlaid on the plot for comparison.}
\end{figure*}

The upper limits on the black hole masses derived in this study provide a unique opportunity to 
extend the $\rm M_{\bullet}-\sigma_{*}$ relation in the low mass regime. The unified relation presented in this work spans a wide range of velocity dispersions 
$(10-300 \kms)$, bridging both the low-mass and extremely high-mass regimes. The sample includes low mass black hole 
measurements from \cite{2005ApJ...634.1093G,2006ApJS..166..249M,2010ApJ...710.1063V,lutzgendorf2013m,2014A&A...566A..58K,2014MNRAS.438..487D} and include ultra massive black holes $(M_{\bullet}>10^{9}M_{\odot})$ measurements from \cite{2002ApJ...578..787C, 2003MNRAS.342..861T,2009ApJ...690..537D,
2011Natur.480..215M,2012ApJ...756..179M,2012MNRAS.419.2497B,2013AJ....146...45R, 2014ApJ...781..112G, 2015MNRAS.452.1792Y,2016ApJ...817....2W,2016Natur.532..340T}. We also add new measurements of ultra-massive black hole masses from \cite{2019ApJ...887..195M,2023MNRAS.521.3298N,2025MNRAS.541.2853M} to the sample. Current ELT/HARMONI simulations \citep{nguyen2023simulating,nguyen2025simulating,ngo2025detecting} rely on simple prescriptions such as the galaxy-black hole scaling relation \citep{2018MNRAS.473.5237K} for assigning black hole masses. The unified $\rm M_{\bullet}-\sigma_{*}$ relation presented in this paper, spanning 7 orders of magnitude, provides a more balanced calibration and offers a stronger empirical foundation for future ELT-based black hole simulations.
In order to produce more complete models of the $M_\bullet -\sigma$ relation across all scales, we have to add the effects of tidal stripping to the black hole growth paradigm of  \cite{bhattacharyya2020cosmic} that  includes the growth of black hole mass and spin through stellar capture, gas accretion, and mergers in a relativistic framework over cosmic time. This will help us improve the mass limits and constraints across a wider mass range which we plan to address in a detailed future study.

We construct an extended and heterogeneous sample of black hole masses by 
assembling measurements from \cite{lutzgendorf2013m, van2016unification, greene2020intermediate}, 
which include black hole masses estimated using stellar and gas dynamics, megamasers, 
and reverberation mapping. We add the upper limits on black hole masses derived in the 
present work from stellar dynamics to this extended sample. Our final sample, therefore, 
contains both direct black hole mass measurements and upper limits. \cite{van2016unification} report an intrinsic scatter equal to $0.49 \pm 0.03 $ in the 
$\rm M_{\bullet}$--$\sigma_{*}$ relation for a sample that includes black hole masses 
estimated using stellar and gas dynamics, megamasers, and reverberation mapping. Similarly, 
\cite{gultekin2009m} constructed a sample using only dynamical methods, excluding black hole 
masses obtained from reverberation mapping. \cite{gultekin2009m} showed that stellar-dynamical and 
megamaser subsample yields a scatter of $0.49\pm0.075$, whereas the gas-dynamical subsample 
yields a somewhat smaller value of $0.35\pm0.096$. They also find that pseudo-bulges exhibit 
an intrinsic scatter of $0.28\pm0.096$ while their full sample has a scatter of 
$0.44\pm0.06$. Taken together, these studies indicate that the intrinsic scatter in the $\rm M_{\bullet}$--$\sigma_{*}$
relation typically lies in the range of $0.3-0.6$ \citep{gultekin2009m, mcconnell2013revisiting}, reflecting the differences 
in black hole mass measurement techniques, galaxy morphology, and fitting methods. We perform Bayesian linear regression using
$emcee$ \citep{foreman2013emcee}, modeling the likelihood for the censored data via the cumulative distribution function 
\citep{kelly2007some,lutzgendorf2013m}. We fit our dataset to 
\begin{equation}
   \log(M_{\bullet})= \alpha + \beta\log(\sigma_{*}/200),
\end{equation}
and find $\alpha=8.32^{+0.04}_{-0.04}$ , $\beta=4.08^{+0.09}_{-0.09}$, with an intrinsic scatter equal 
to $\sigma_{int}=0.55^{+0.03}_{-0.02}$. We show the unified $\rm M_{\bullet}$--$\sigma_{*}$ relation spanning stellar 
velocity dispersions from $\sim 10~\mathrm{km\,s^{-1}}$ to $\sim 300~\mathrm{km\,s^{-1}}$ in Figure 4. We find that the slope of  
the $\rm M_{\bullet}$--$\sigma_{*}$ relation $\beta = 4.08$ lies within the typical range of $\beta \sim 3.68$ – $5.35$ reported for a diverse population of galaxies (e.g., \cite{ferrarese2000fundamental, gebhardt2000relationship, merritt2001black, tremaine2002slope, graham2008fundamental, gultekin2009m, van2016unification, bhattacharyya2018m_}). We find an intrinsic scatter of $\sigma_{int} = 0.55^{+0.03}_{-0.02}$ in the $M_{\bullet}$-$\sigma_{*}$ relation, compared to $\sigma_{int} = 0.44 \pm 0.06$ \citep{gultekin2009m} and $\sigma_{int} = 0.49 \pm 0.03$ \citep{van2016unification}. The increased scatter in our sample could be due to the inclusion of low-mass galaxies and upper limits in the fit.

The $\rm M_{\bullet}$--$\sigma_{*}$ relation presented in this work indicates that dwarf spheroidal galaxies can host black holes with masses in the range of $10^{2}$--$10^{3}\, M_{\odot}$, consistent with a scenario where black holes accrete mass by exchanging momentum with the surrounding medium. 
Whereas, a black hole can grow through stellar capture up to $10^{4}\, M_{\odot}$, even when the accretion-driven growth has stalled. The black hole masses predicted from both stellar capture and gas accretion lie within the upper limits obtained in our study. \cite {pacucci2023extreme}, present an alternative scenario for presence of massive black holes $\sim 10^{6}M_{\odot}$ in dwarf spheroidal galaxies. \cite {pacucci2023extreme} suggests that the progenitors of present-day dwarf spheroidal galaxies, initially hosting stellar masses of $\sim 10^6\,M_\odot$, may have occupied higher positions 
on the \( M_{\bullet}\!-\!\sigma_* \) relation. However, mass loss of up to \(\sim58\%\) during pericentric passages can significantly reduce their 
stellar content, shifting them to the lower end of the observed \( M_{\bullet}\!-\!\sigma_* \) relation, while leaving the central black hole mass unchanged. Tidal stripping of the progenitors of present-day dwarf spheroidal galaxies may be an important channel through which these systems can host black hole masses comparable to the upper limits obtained in our study. However, black hole growth via momentum-driven accretion and stellar capture yields more conservative mass estimates, broadly consistent with the predictions of the \( \rm M_{\bullet}\!-\!\sigma_{*} \) relation.

\section{Conclusions}
In this work, we have constructed dynamical models of dwarf spheroidal galaxies to constrain the black hole mass in conjunction with the parameters corresponding to the dark matter halo using anisotropic distribution functions. We find that the posterior for the dark matter mass is strongly constrained with a median value between $10^{8} - 10^{9}M_{\odot}$. On the other hand, the posterior distribution for the black hole mass remains flat towards the low mass end. We derive \(95\%\) credible upper limits on the black hole masses in dwarf spheroidal galaxies, which lie in the range \(10^{5} - 10^{6}\,M_{\odot}\). We combine the upper limits obtained in our study for dwarf spheroidal galaxies with previously reported black hole mass measurements and upper limits from the literature to derive a unified \( M_{\bullet}\!-\!\sigma_{*} \) relation spanning $\sigma_{*}\sim 10\kms - 300\kms$.We find 
that the \( \rm M_{\bullet}\!-\!\sigma_{*} \) relation is given by $\log(M_{\bullet}) = 8.32 + 4.08\,\log\left(\sigma_{*}/200\,\mathrm{km\,s^{-1}}\right),$
with an intrinsic scatter of \( \sigma_{\rm int} = 0.55\). We compare different formation scenarios through which the dwarf spheroidal galaxies can host central black holes consistent with the upper limits derived in our study. The upper limits on black hole masses obtained in our work (\( \leq 10^{6}\,M_{\odot} \)) are consistent with a scenario in which the progenitors of present-day dwarf spheroidal galaxies experienced tidal stripping, reducing their stellar mass while leaving the central black hole largely unaffected. We conclude that models of black hole growth via momentum-driven gas accretion and stellar capture yield black hole masses between $10^{3}-10^{4}M_{\odot}$, closer to the prediction of $\rm M_{\bullet}$--$\sigma_{*}$ relation within the 95$\%$ credible upper limits on the black hole masses presented in this work.

\section{Acknowledgments}
The authors thank the referee for the insightful comments and questions, which have greatly helped in improving the manuscript.
We gratefully acknowledge the use of the high-performance computing facility 'NOVA' at the Indian Institute of Astrophysics, Bengaluru, India, where all computations were carried out. The work is supported by ANRF grant CRG/2021/005174/PHY.

\newpage
\bibliography{2366_d2.bib}{}
\bibliographystyle{aasjournalv7.bst}



\end{document}